# A LABORATORY STUDY OF THE REDUCTION OF IRON OXIDES BY HYDROGEN

Damien Wagner, Olivier Devisme, Fabrice Patisson, Denis Ablitzer

Laboratoire de Science et Génie des Matériaux et de Métallurgie (LSG2M)
UMR 7584 CNRS-INPL-UHP, Nancy School of Mines
Parc de Saurupt, 54042, Nancy Cedex, France

## Abstract

To reduce the emission of greenhouse gases by the steel industry, particularly for ironmaking, the production of DRI (Direct Reduced Iron) using hydrogen as the reducing gas instead of carbon monoxide is being considered. In this context, the reduction of pure hematite by hydrogen was studied at the laboratory scale, varying the experimental conditions and observing the rate and the course of the reaction. All the reduction experiments were performed in a thermobalance and supplementary characterization methods were used like scanning and transmission electron microscopy, X-ray diffraction, and Mössbauer spectrometry. The influence of rising temperature in the range 550-900°C is to accelerate the reaction; no slowing down was observed, contrary to some literature conclusions. A series of experiments consisted in interrupting the runs before complete conversion, thus enabling the characterization of partially reduced samples. Interpretation confirms the occurrence of three successive and rather separate reduction steps, through magnetite and wustite to iron, and illustrates a clear structural evolution of the samples. Finally, the influence of the sample type was revealed comparing a regular powder, a nanopowder and a sintered sample. The regular powder proved to be the most reactive despite its larger grain size, due to a more porous final structure.

## Introduction

A European consortium of steel producers led by Arcelor has launched a research program called ULCOS (Ultra Low $CO_2$ Steelmaking) with the objective of investigating new steel production processes to obtain a 50% reduction in the total greenhouse gases (GHG) emissions from the steel industry [1]. The program involves 48 industry and university partners and is sponsored by the European Commission. One of the processes under study is the direct reduction of iron ore in a shaft furnace using pure hydrogen to produce DRI (Direct Reduced Iron). The work presented here is part of LSG2M's task in the ULCOS project. It consisted in a series of laboratory experiments aiming at observing the course of the reduction of hematite by pure hydrogen and the influence of temperature and initial morphology on the reaction rate. Another part of our work, that will not be dealt with here, is to develop a 2-D mathematical model of the hydrogen-based shaft furnace process.

The reduction of iron ores by hydrogen is a gas-solid reaction which occurs in two or three stages. For temperatures higher than 570°C, hematite ($Fe_2O_3$) is first transformed into magnetite ($Fe_3O_4$), then into wustite ($Fe_{1-y}O$), and finally into metallic iron whereas at temperatures below 570°C, magnetite is directly transformed into iron since wustite is not thermodynamically stable (see Figure 1).

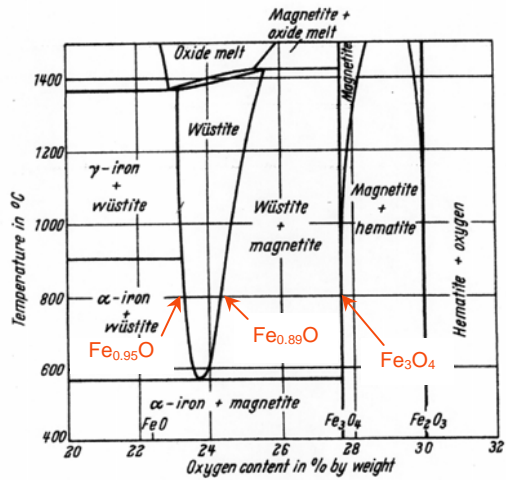

Figure 1. Stability diagram of the different iron oxides as
a function of temperature and oxygen content [2].

The reduction of iron ore by hydrogen was widely studied in the 60s to 80s. Most of the reaction features are very similar to that of the reduction by carbon monoxide and many mechanisms are common to both of them [3-7]. Nevertheless, some significant differences must be underlined. First of all, the reduction with hydrogen is endothermic, whereas it is exothermic with carbon monoxide. Conversely, thermodynamics are more favorable with hydrogen than with carbon monoxide above 800°C. This makes the industrial operation different. With hydrogen, the hot gas fed has to bring enough calories to heat and maintain the solid at a temperature sufficiently high for the reaction to occur. Operating with a gas flowrate higher than stoichiometry is therefore necessary. Kinetics are also reported to be faster with hydrogen. This in turn can modify the morphology of the final product (iron), which depends on a competition between diffusion and chemical reaction [2, 6-8]. In particular, the formation of whiskers seems a specific feature of the reduction by hydrogen [3]. Whiskers are iron grains protruding from the wustite phase and growing as fingers toward the exterior of the particles. According to the literature [9], they make the iron-iron contacts more frequent and could thus explain the phenomenon of sticking of the solid particles, sometimes experienced in industrial reactors operated with a high hydrogen content. Another awkward phenomenon observed with hydrogen is the occurrence, at some temperatures, of a slowing down at the end of the reaction to reach the last percents of conversion degree. Before determining the best operating conditions for performing the reaction industrially, it seemed to us necessary to undertake a new laboratory study of the reaction of iron oxide reduction by hydrogen. The present paper reports our first findings.

**Experimental Apparatus**

We carried out the reduction experiments using a thermobalance. Recording the evolution of the sample mass as the transformation occurs enables to follow the different steps of the reaction and to obtain kinetic data for each of them. The thermobalance used was a SETARAM TAG 24 whose particularity is to have two symmetrical furnaces, one for the sample (sample in crucible) and one for the reference (empty crucible). This configuration gives a more accurate weight measurement than that of usual, single furnace apparatus. Indeed, buoyancy and drag forces, acting the same way on the two crucibles, vanish instead of interfering with weight in the case of a single furnace. A schematic representation of the experimental device is presented in Figure 2. The gases used were pure hydrogen, argon and helium (Air Liquide, grade Alphagaz 1). Part of the argon was sent to the resistor to protect it, the other gases being the carrier gases (argon and helium) and the reacting gas (hydrogen). Each flowrate could be modified to obtain different hydrogen contents in the gaseous mixture. The flowrate of the gas passing through the sample furnace was obtained by substraction between the total gas flowrate measured at the head of the balance, and that of the gas flowing through the reference furnace measured at its outlet. A needle valve placed at the outlet of the sample furnace was used to ensure that there was the same gas flowrate passing through the two furnaces. For all of the experiments, the gas flowrate was set to 70 cm$^3_{STP}$/min in each furnace.

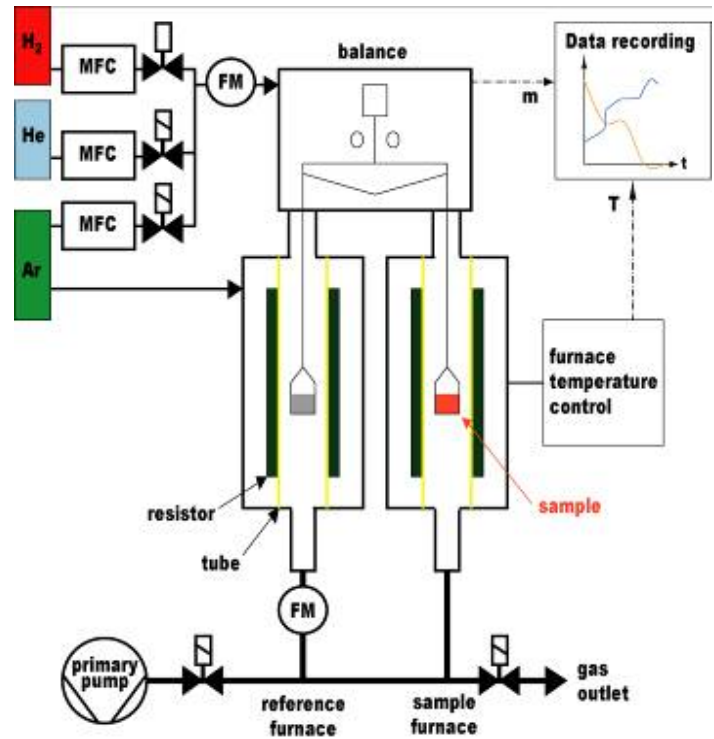

Figure 2. Schematic representation of the experimental device.

Solid samples used were three synthetic hematites, manufactured by Aldrich. The main type was a 99.9%-pure hematite powder, with grains smaller than 5 µm. It will further be referenced to as P1. The two other sample types were small sintered pieces, with grains about 1 µm in diameter (S1), and a very fine "nanopowder" (N1).

**Experimental Procedure**

The hematite sample (usually of powder P1) was first charged in an alumina crucible, weighed on a Mettler balance with a $10^{-5}$ g precision, then installed in the thermobalance. The balance was emptied, then filled with helium and a constant flow of this gas was maintained during 30 to 60 minutes in order to sweep the furnaces and to adjust the gas repartition. The two furnaces were then heated at 50 K/min to the temperature of the experiment and the data collection was started. After waiting for 10 minutes to stabilize temperature, the hydrogen valve was opened and the reduction proceeded. When the time chosen for the end of the experiment was reached, the valve was closed and a large amount of argon was blown into the balance so as to expel the hydrogen. The reaction was thus stopped and the balance was let under a helium flow until getting back to room temperature.

A series of interrupted experiments was carried out in order to clarify the course of the reaction. As the reaction was too fast when using pure hydrogen, a mixture with 10 % in helium was used. The total conversion time was thus increased from 350 seconds to approximately 1100 seconds at 800°C. Figure 3 presents the evolution of the fractional mass loss and reduction degree as a function of time.

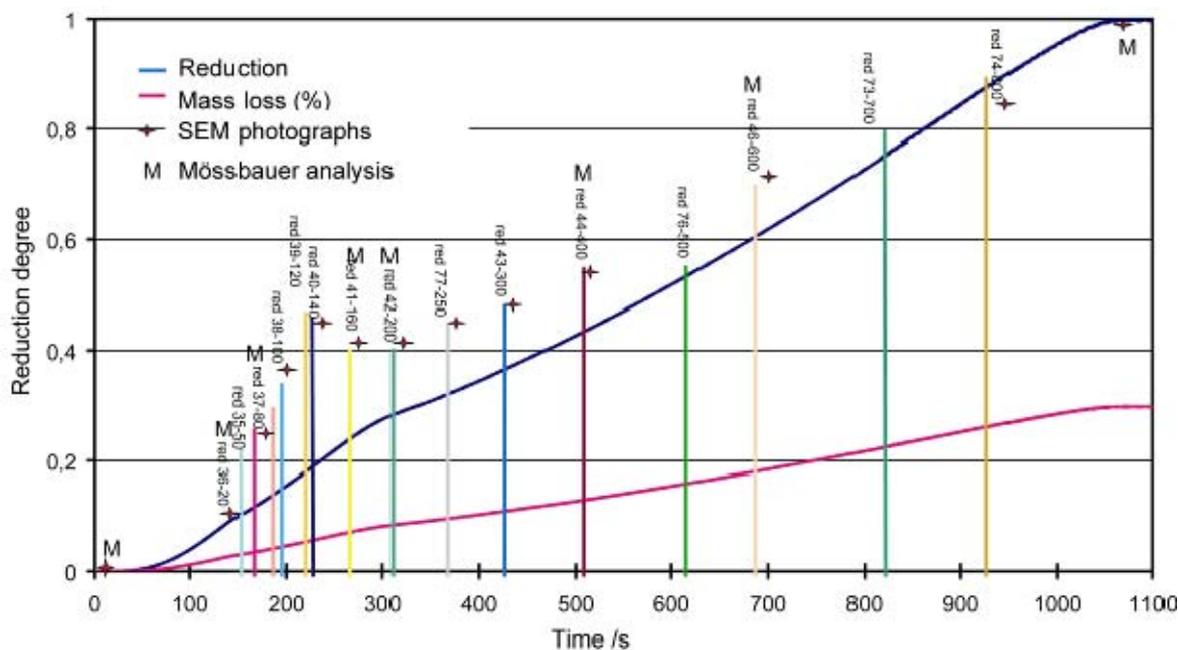

Figure 3. Evolution of the reduction degree and distribution of the interrupted experiments.

Several changes in the slope can be noticed on the curves. Each of them, as it will be shown, can be attributed to the onset of a new reaction corresponding to the different iron oxides. The interrupted experiments were distributed along the curve so as to obtain data about the different steps of the transformation. The distribution is indicated in Figure 3. Each vertical line corresponds to an interrupted experiment and a X-ray diffraction analysis. For some of them, a Mössbauer quantitative analysis was also carried out and/or scanning electron microscopy (SEM) photographs were taken.

Other experiments were performed to investigate the influence of temperature between 550 and 900°C and of the sample type, using the nanopowder and the sintered pieces. The experimental procedure was always the same as the one presented above.

**Results and Discussion**

Interrupted Experiments

Each sample was characterized by X-ray diffraction. All of the spectra obtained were superposed in Figure 4. It shows that the reduction of hematite into magnetite is very fast since as soon as Red 35 sample (167 s), no hematite peak can be observed any more (frame H). The sample appears only composed of magnetite. On the following sample Red 37 (189 s), a new oxide is detected whose peaks correspond to wustite. According to the frames Wa and Wb, wustite is present in the samples for almost the entire time of the reduction, its peaks increase in size as the magnetite ones reduce. In Red 43, the first iron particles appear. From this sample, the iron peak (frame F) increases as the wustite peaks shift a little and decrease in size. The movement of these peaks could come from the evolution of the stoichiometry from $Fe_{0.89}O$ to $Fe_{0.95}O$. The last sample Red 75 is composed of pure iron.

These results suggest that the different steps involved in the reduction of hematite to iron are quite well separated since wustite only appears when there is no more hematite and iron appears when there is no more magnetite.

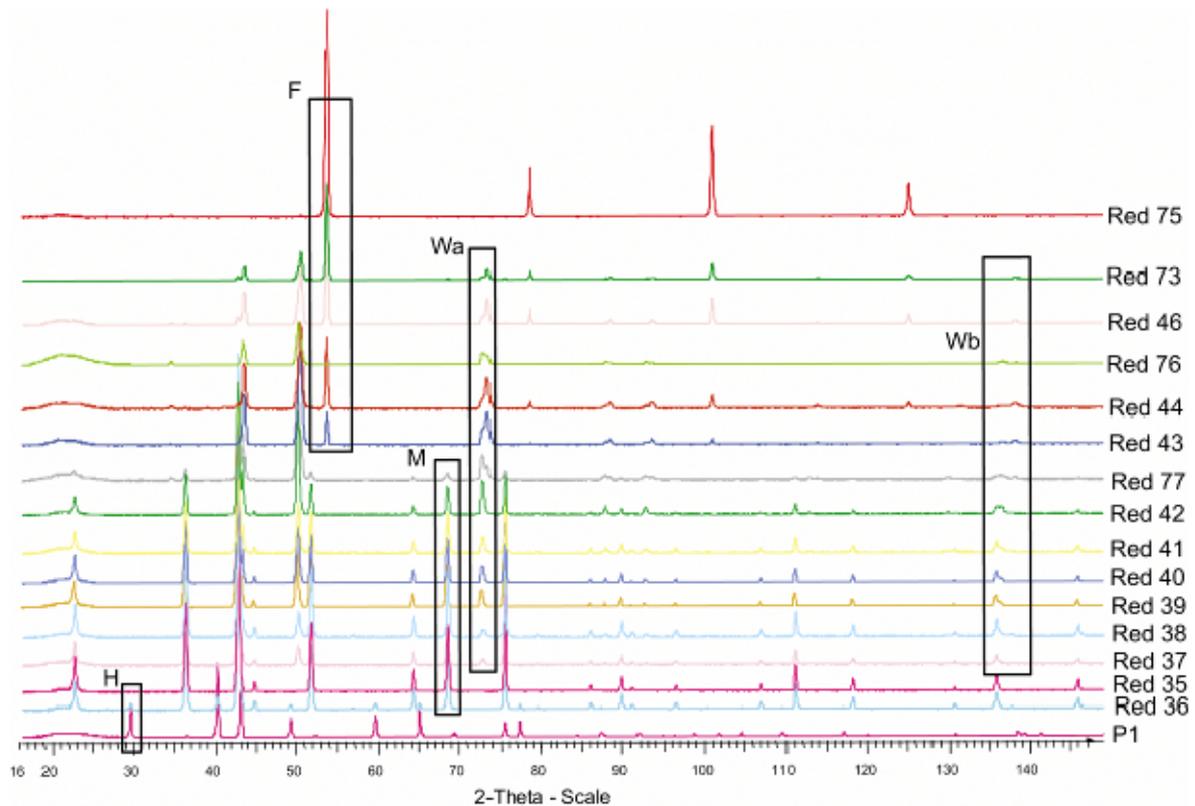

Figure 4. X-Ray spectra for the reduction at 800°C and 10 % of $H_2$.

To confirm the previous results, Mössbauer analyses were performed. Figure 5 shows the evolution of the composition of the partially reduced samples as a function of time, as well as the complete reduction curve given by thermogravimetry. The theoretical positions of the intermediate iron oxides are also reported on the right axis.

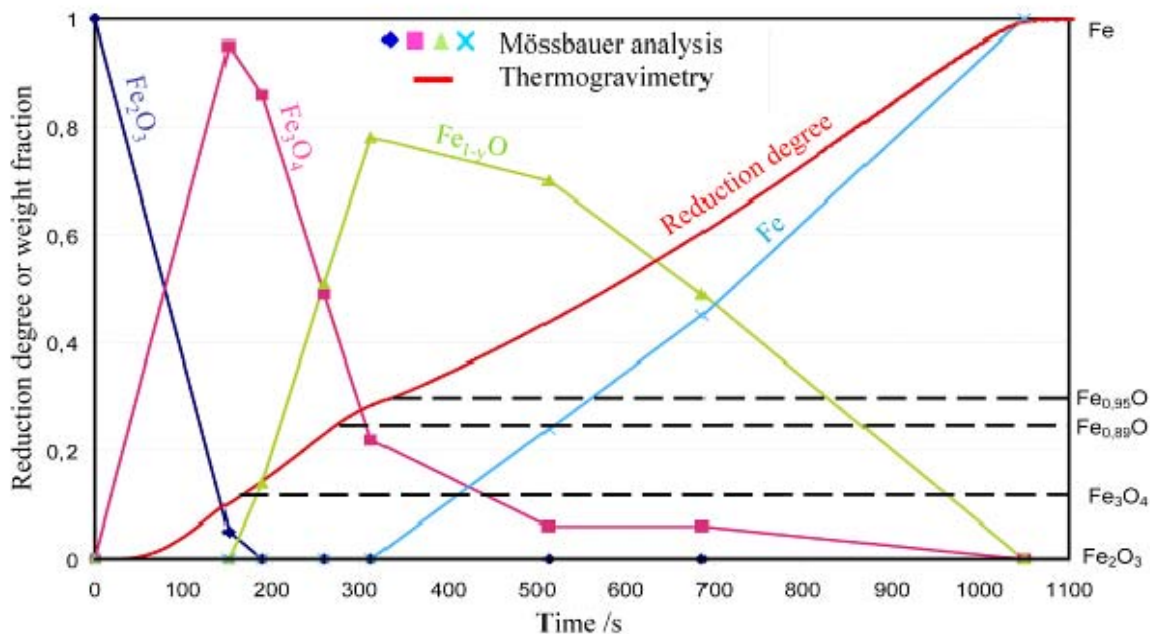

Figure 5. Results of the Mössbauer analyses (reduction at 800°C, 10 % $H_2$).

The formation of each intermediate oxide to the detriment of the previous one can clearly be observed. The transformation from hematite to magnetite is fast and complete. For the second step, wustite appears as magnetite is consumed but wustite begins to be reduced while magnetite remains until the end. This late presence of magnetite contradicts the X-ray diffraction results and can be attributed to a lower sensitivity of the X-ray analysis. The reduction of wustite is the longest step of the whole reduction process. On the conversion curve, it can be noticed that all of the changes in slope roughly correspond to the theoretical positions of the oxides and also to the occurrence of the maxima in magnetite (~150 s) and wustite (~300 s).

Figure 6 presents SEM photographs of samples of the solid obtained after the interrupted experiments. The initial hematite is composed of small aggregates of approximately 1 µm, which are themselves made of smaller grains. At a higher magnification, these grains appear to be dense and of irregular shapes. The porosity between the particles is high. From the beginning of the reduction to Red 37, which corresponds to the transformation of hematite into magnetite, there are little morphological changes. One can only notice a somewhat more continuous structure of the solid, with the smaller constituting grains less visible.

Between Red 37 and Red 43, which corresponds to the conversion of magnetite into wustite, the changes are more obvious, though it is difficult to distinguish between magnetite and wustite phases. For example, according to the Mössbauer analyses, Red 37 is mainly composed of magnetite and Red 43 of wustite whereas Red 41 is likely made of 50 % of each. The final structure of wustite exhibits larger, flatter particles, more like discs than like spheroids. Most of the morphological changes occur at the beginning of the transformation and then, nothing visible happens. This suggests that wustite covers the surface of the particles very quickly, gathers neighbor grains and that further transformation takes place in the interior of the grains.

Finally, the reduction from wustite to iron (Red 43 to 75) entails dramatic changes in the morphology of the particles, as can be seen on the three last photographs. The iron phase is clearly identifiable. It appears as larger-size (about 2 µm) particles, often finger-like shaped. The growth seems to occur toward the exterior of the particles and the grains often join to give bridges. On the Red 75 photograph, the iron phase is quite continuous, but porous, with grains of approximately 3 µm. The small bright points at the surface are the last wustite grains that were reduced. This kind of growth was already observed and explained by many authors [6, 9-12]. It results from a competition between the chemical reaction at the surface of the particles, which increases the iron concentration and activity at this place, and the volumetric diffusion of this excess iron toward the core of the particle. An external, whisker-type growth, is caused by a reaction chemically controlled, whereas a growth in the form of a spread layer is due to a diffusion control.

Influence of Some Experimental Parameters

The results of the experiments carried out at different temperatures are gathered in Figure 7. A global acceleration of the reaction rate with temperature can be noticed. On the contrary to some authors [13, 14], no slowing down could be observed between 700 and 800°C. As expected, no second change in slope can be seen on the curve at 550°C since wustite is not stable at this temperature.

The reaction rates at 5 % and 80 % of conversion, corresponding respectively to the transformation from hematite to magnetite and from wustite to iron, were reported on an Arrhenius diagram. The activation energies calculated are 14.7 kJ/mol and 14.3 kJ/mol. These values are low compared to most of those reported in the literature [15-18]. An explanation could be the high porosity of our powder samples (contrary to compact and dense particles usually studied) that favors a gas-diffusion control. However, this point requires further investigations.

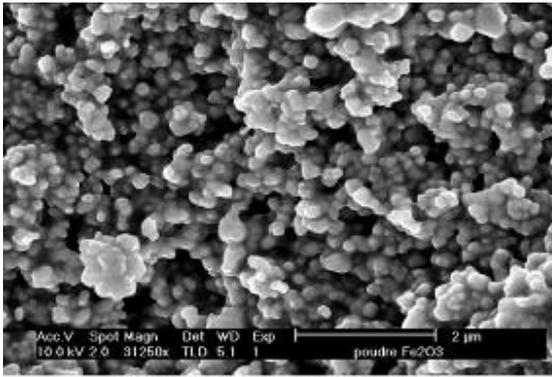
Initial hematite powder P1

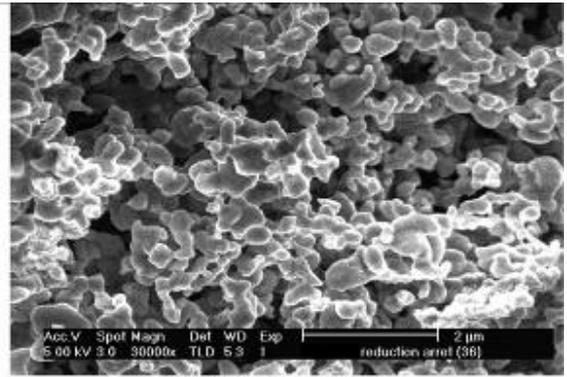
Red 36, after 152 s

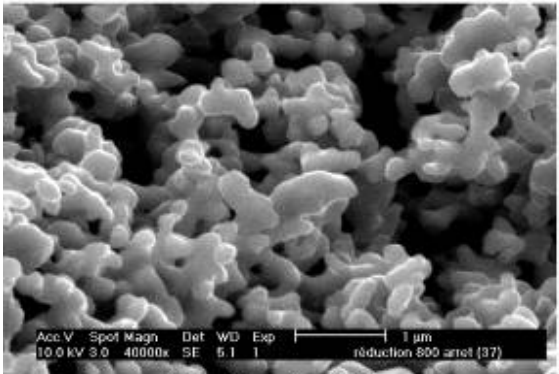
Red 37, after 189 s

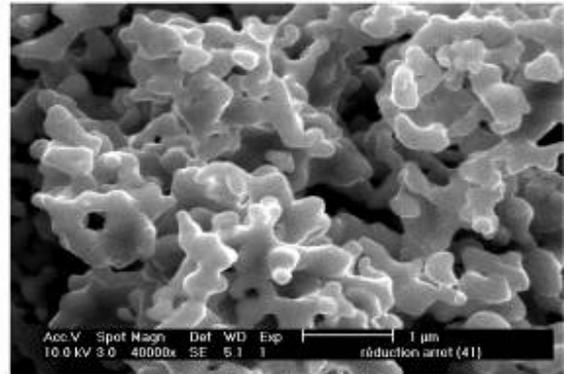
Red 41, after 260 s

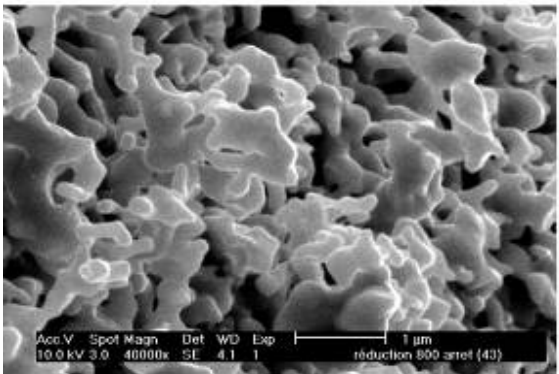
Red 43, after 427 s

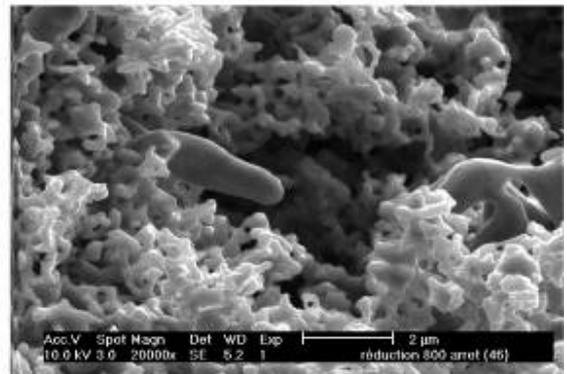
Red 46, after 686 s

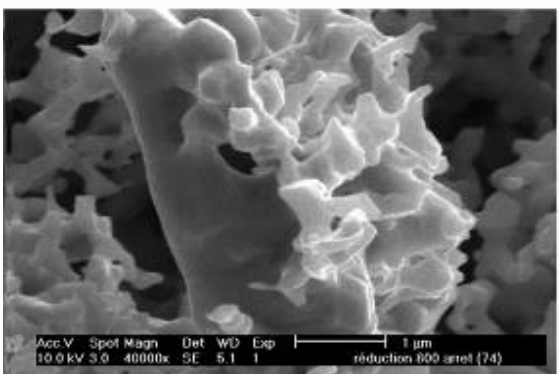
Red 74, after 935 s

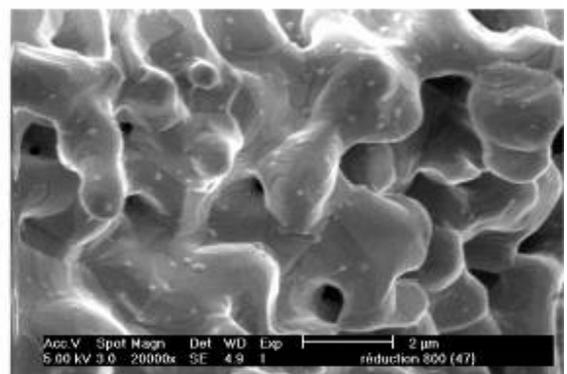
Red75, final iron

Figure 6. SEM phographs for different reduction times

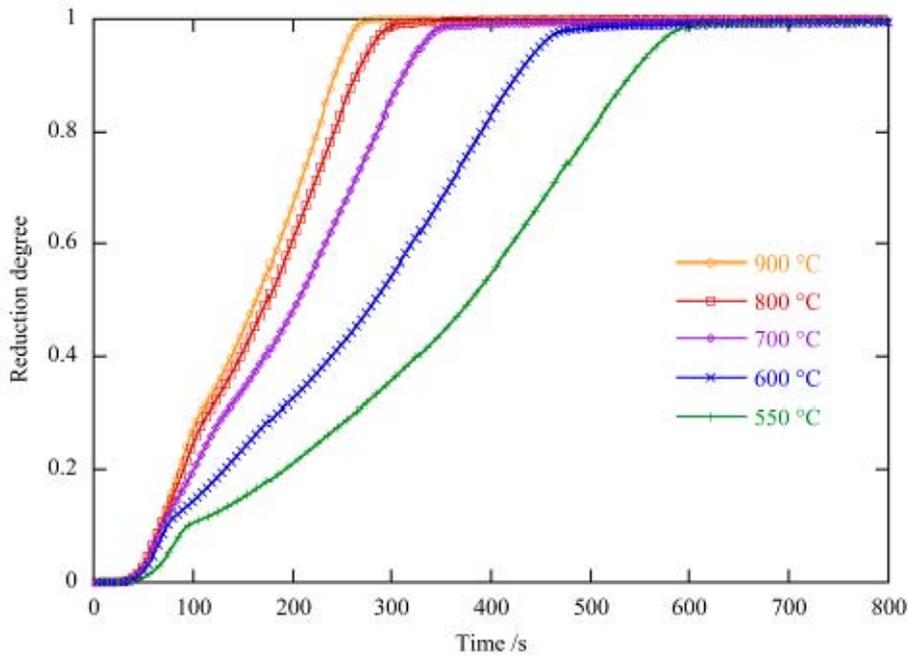

Figure 7. Influence of temperature on the reaction rate (100 % $H_2$).

Some experiments using the two other oxides (S1 and N1) were performed in order to investigate the influence of the sample type. The kinetics are reported in Figure 8. The reaction is faster for the coarse powder (P1) whereas the two other solids exhibit a similar behavior. This result is *a priori* surprising since the specific area is higher for N1 and S1 than for P1.

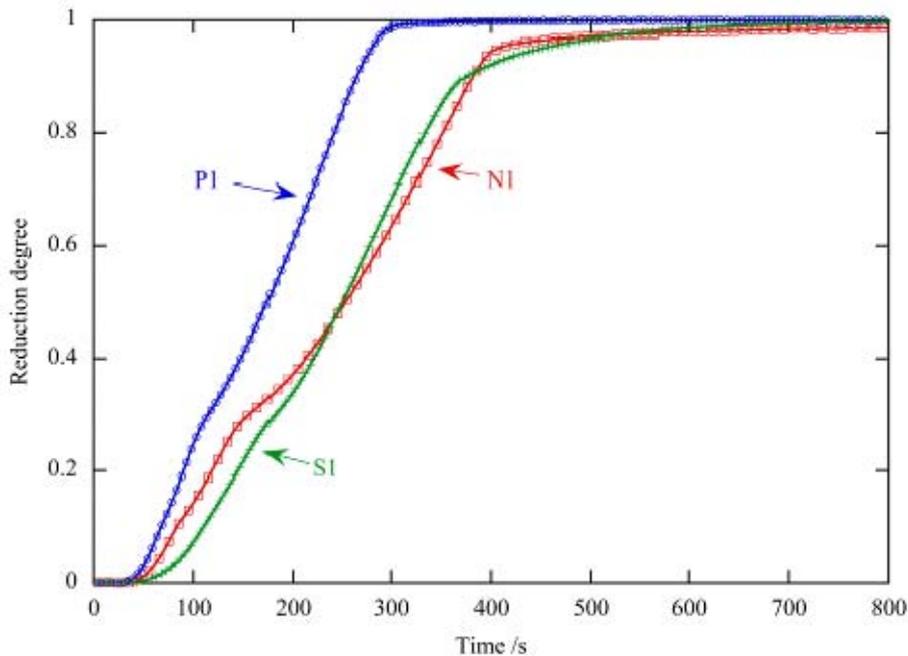

Figure 8. Influence of the sample type on the reduction of hematite at 800°C with pure $H_2$.

So as to understand the reasons of this behavior, cross-sections of the samples before and after the reduction were made. These are presented in Figure 9. For the coarse sample P1, there has been a change in the morphology of the particles, already discussed, but the porosity remains quite high allowing easy gaseous diffusion. On the contrary, for S1 and N1, whereas the initial state also presents a great porosity, the final product is much more compact making the limiting step to become solid-state diffusion, which is far slower.

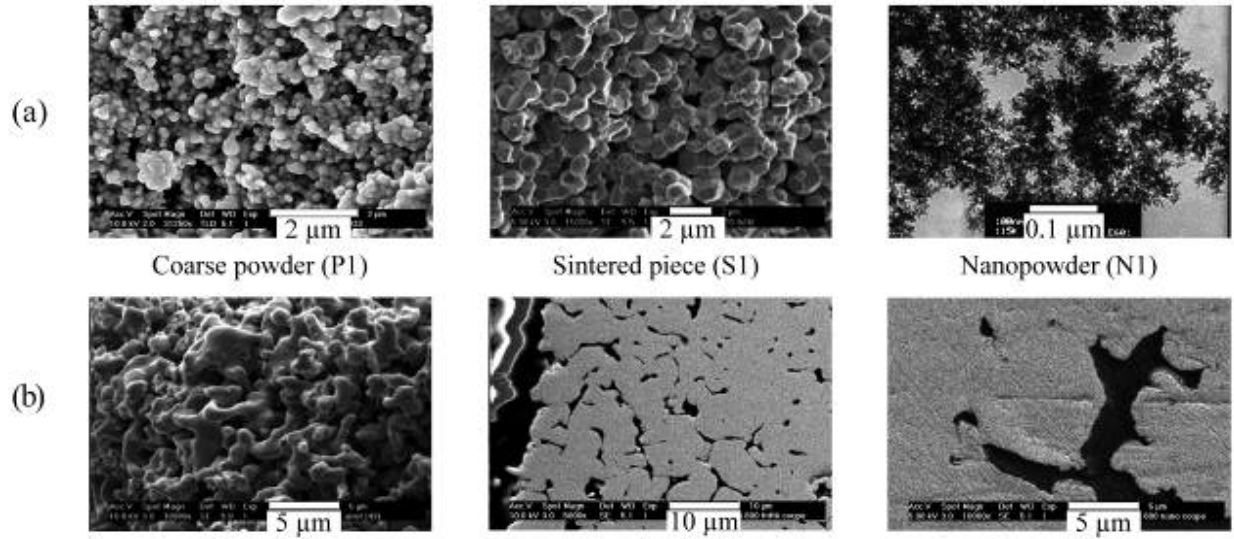
Figure 9. Cross sections of the different samples before (a) and after (b) reduction.

**Conclusion**

The reduction of synthetic hematite samples is a multi-stage reaction with one or two intermediate oxides depending on temperature. Thermogravimetric experiments and analyses of partially reduced samples have shown that the longest step is the last transformation, i.e. from wustite to iron. The first two reactions, hematite to magnetite and magnetite to wustite, are successive and well separated since hematite has completely disappeared when the first grains of wustite are detected. On the contrary, the reduction of wustite into metallic iron begins before the total consumption of magnetite.

The morphological changes, barely observed in transformation of hematite into magnetite, are significant in the last two reactions. The initial small spherical grains first turn into flatter wustite grains and finally into more massive, finger-like shaped, connected iron particles. The iron phase seems to grow outwards. This kind of morphology could lead to the sticking phenomenon.

Besides, the influence of temperature and sample type were studied. In the experimental conditions considered, and in the range 550-900°C, an increase in temperature accelerates the reaction. Experiments with three types of hematite samples have shown differences in reactivity. With the sintered pieces and the nanopowder, the final iron structure can be quite dense, making the gaseous diffusion very difficult. The solid state diffusion thus probably becomes the limiting step of the kinetics and the reaction rate is lowered even if the initial specific area of the sample was high.

**Acknowledgement**

We thank P. Delcroix for his assistance in Mössbauer analyses and Drs. J.P. Birat and E. Hess from Arcelor Research for helpful discussions on this subject. We also acknowledge the financial support of European Union in the frame of the ULCOS project (FP6, N° 515960).

**References**

1.     J.P. Birat, F. Hanrot, and G. Danloy, "CO$_2$ Mitigation Technologies in the Steel Industry: a Benchmarking Study Based on Process Calculations", *Stahl und Eisen*, 123 (9) (2003), 69-72

2. L. V. Bogdandy and H. J. Engell, *The Reduction of Iron Ores* (Berlin, Springer-Verlag, 1971)

3. R. Nicolle and A. Rist, "Mechanism of Whisker Growth in the Reduction of Wustite", *Metallurgical Transactions B-Process Metallurgy*, 10 (3) (1979), 429-438

4. C. Wagner, "Mechanism of the reduction of oxides and sulfides to metals", *Journal of Metals*, 4 (Trans.) (1952), 214-6

5. D. H. St John, S. P. Matthew and P. C. Hayes, "The Breakdown of Dense Iron Layers on Wustite in CO/CO2 and H2/H2O Systems", *Metallurgical Transactions B-Process Metallurgy*, 15 (4) (1984), 701-708

6. D. H. St John, S. P. Matthew and P. C. Hayes, "Establishment of Product Morphology During the Initial-Stages of Wustite Reduction", *Metallurgical Transactions B-Process Metallurgy*, 15 (4) (1984), 709-717

7. D. H. St John and P. C. Hayes, "Microstructural Features Produced by the Reduction of Wustite in H2/H2O Gas-Mixtures", *Metallurgical Transactions B-Process Metallurgy*, 13 (1) (1982), 117-124

8. D. H. St John, S. P. Matthew and P. C. Hayes, "Microstructural changes during the reduction of wustite", *Symposia Series - Australasian Institute of Mining and Metallurgy*, 36 (Extr. Metall. Symp.) (1984), 287-94

9. J. F. Gransden and J. S. Sheasby, "Sticking of Iron-Ore During Reduction by Hydrogen in a Fluidized-Bed", *Canadian Metallurgical Quarterly*, 13 (4) (1974), 649-657

10. S. Elmoujahid and A. Rist, "The Nucleation of Iron on Dense Wustite - a Morphological-Study", *Metallurgical Transactions B-Process Metallurgy*, 19 (5) (1988), 787-802

11. I. Gaballah, P. Bert, L. C. Dufour and C. Gleitzer, "Kinetics of the reduction of wustite by hydrogen and carbon monoxide + hydrogen mixtures. Observation of trichites", *Mémoires Scientifiques de la Revue de Métallurgie*, 69 (7-8) (1972), 523-30

12. J. F. Gransden, J. S. Sheasby and M. A. Bergougnou, "Defluidization of iron ore during reduction by hydrogen in a fluidized bed", *Chemical Engineering Progress, Symposium Series*, 66 (105) (1970), 208-14

13. M. Moukassi, P. Steinmetz, B. Dupre and C. Gleitzer, "Mechanism of reduction with hydrogen of pure wustite single crystals", *Metallurgical Transactions B: Process Metallurgy*, 14B (1) (1983), 125-32

14. E. T. Turkdogan and J. V. Vinters, "Gaseous Reduction of Iron Oxides.1. Reduction of Hematite in Hydrogen", *Metallurgical Transactions*, 2 (11) (1971), 3175-3188

15. W. Pluschkell and H. Yoshikoshai, "Growth of the iron phase on wustite during reduction in mixtures of hydrogen and water vapor", *Archiv fuer das Eisenhuettenwesen*, 41 (8) (1970), 715-21

16. E. T. Turkdogan and J. V. Vinters, "Gaseous Reduction of Iron Oxides.3. Reduction-Oxidation of Porous and Dense Iron Oxides and Iron", *Metallurgical Transactions*, 3 (6) (1972), 1561-1574

17. S. K. El-Rahaiby and Y. K. Rao, "The kinetics of reduction of iron oxides at moderate temperatures", *Metallurgical Transactions B: Process Metallurgy*, 10B (2) (1979), 257-69

18. P. C. Hayes, "The kinetics of formation of water and carbon dioxide during iron oxide reduction", *Metallurgical Transactions B: Process Metallurgy*, 10B (2) (1979), 211-17